\documentclass[9pt,twocolumn,twoside]{pnas-new}
\usepackage{bm}
\usepackage{bbm}
\templatetype{pnasresearcharticle}

\title{Simplicial persistence of financial markets: \\
filtering, generative processes and portfolio risk}

\author[a,1]{Jeremy D. Turiel}
\author[a]{Paolo Barucca}
\author[a,b]{Tomaso Aste} 

\affil[a]{Department of Computer Science, UCL, Gower Street, WC1E6BT London, UK}
\affil[b]{Systemic Risk Centre, London School of Economics and Political Sciences, London, United Kingdom}
\affil[1]{Corresponding author. E-mail: jeremy.turiel@gmail.com}

\leadauthor{Turiel}

\authorcontributions{Please provide details of author contributions here.}
\authordeclaration{Please declare any conflict of interest here.}
\equalauthors{\textsuperscript{1}A.O.(Author One) and A.T. (Author Two) contributed equally to this work (remove if not applicable).}

\correspondingauthor{\textsuperscript{1}To whom correspondence should be addressed. E-mail: jeremy.turiel@gmail.com}
\begin{abstract}
We introduce simplicial persistence, a measure of time evolution of network motifs in subsequent temporal layers. 
We observe long memory  in the evolution of  structures from correlation filtering, with a two regime power law decay in the number of persistent simplicial complexes. 
Null models of the underlying time series are tested to investigate properties of the generative process and its evolutional constraints. 
Networks are generated with both TMFG filtering technique and thresholding showing that embedding-based filtering methods (TMFG) are able to identify higher order structures throughout the market sample, where thresholding methods fail.
The decay exponents of these long memory processes are used to characterise financial markets based on their stage of development and liquidity. 
We find that more liquid markets tend to have a slower persistence decay. This is in contrast with the common understanding that developed markets are more random. 
We find that they are indeed less predictable for what concerns the dynamics of each single variable but they are more predictable for what concerns the collective evolution of the variables.  
This could imply higher fragility to systemic shocks.
\end{abstract}


\dates{\today}

\begin{document}

\maketitle

\ifthenelse{\boolean{shortarticle}}{\ifthenelse{\boolean{singlecolumn}}{\abscontentformatted}{\abscontent}}{}

\section{\label{intro}Introduction}

Networks representing the structure of the interactions within complex systems have been increasingly studied in the last few decades \cite{newman2018networks}. 
Applications range from biological networks to social networks, infrastructures and finance \cite{strogatz2001exploring}. 
In finance - mainly due to the abundance of time-series data regarding economic entities and the lack of data on direct relationships between them - there has been an extensive focus on the estimation of pairwise interactions from pair correlations ~\cite{mantegna1999hierarchical} of the stochastic time series characterising financial markets. 
The need to extract significant links from noisy correlation matrices has triggered the development of filtering techniques which yield sparse network structures ~\cite{jovanovic2018financial, cimini2019statistical, kojaku2019constructing, masuda2018configuration} based on a limited set of statistical or topological hypotheses.
There are three main approaches to network filtering: thresholding, statistical validation, and topological filtering. 
These methods show that a meaningful and consistent taxonomy of financial assets emerges from sparse network structures, in particular when applying topological methods. 
Thresholding methods remove edges which are less significant based on their strength (or its absolute value), quantile thresholding is one of these methods that we use in this paper. 
This method considers the distribution of edge strengths and removes edges with strength below a certain quantile level. 
It is often applied to financial correlation matrices due to its lack of assumptions on the underlying distribution.
Statistical validation - which constitutes a generalisation of simpler thresholding methods - has been used to establish the significance of edges in correlation matrices, with applications to economics and finance as well as other fields \cite{musciotto2018bootstrap, tumminello2011statistically, micciche2019primer, marcaccioli2019polya}.
Statistical validation can be implemented by comparing empirical networks with random networks from constrained randomisations which generate weighted ensembles of null models and allow to quantify the significance of observed realisations with respect to the ensemble statistics of the null constrained model.
Topological filtering through the Minimum Spanning Tree (MST) technique was initially suggested by Mantegna ~\cite{mantegna1999hierarchical}, and was further extended to planar graphs with the Planar Maximally Filtered Graph (PMFG) ~\cite{tumminello2005tool} and more recently to chordal graphs with predefined motif structure, as the Triangulated Maximally Filtered Graphs (TMFG) in ~\cite{massara2016network} and the Maximally Filtered Clique Forest (MCFC) in ~\cite{massara2019learning}.

%The development of null models and the study of their applications constitutes an important discipline within statistical physics. 
%These methods aim for a constrained randomisation of the considered system which allows to generate weighted ensembles and study the significance of observed realisations with respect to the ensemble. 
%Constraints often imply as underlying generative process and can help understanding if the assumptions are correct as well as how and where observed realisations deviate from these assumptions. 

Market efficiency imposes the absence of temporal memory in log returns, but the presence of long memory in higher-order moments of returns and long-term dependence (autocorrelation) of absolute and squared returns have been observed and the are now considered among the important stylised facts in markets \cite{cont2001empirical}, e.g. volatility clustering, a form of regime switching in the fluctuations observed in financial markets. 
In \cite{lillo2004long, PhysRevE.71.066122}, later extended in \cite{ bouchaud2009markets}, it was shown that order signs obey a long memory process, balanced by anti-correlated volumes which guarantee market efficiency.
In financial time series analysis, through the generalised Hurst exponent analysis, it was demonstrated that memory effects are related to the stage of maturity of the market, with more mature markets being more random \cite{di2005long}.

With the present paper we provide the missing piece, connecting market structure and market memory by analysing the autocorrelation of market structures \cite{PhysRevE.88.012806}, through persistence of its filtered correlation matrix \cite{PhysRevE.86.026101}. 
We analyse a range of null models \cite{PhysRevE.68.046130} - corresponding to a range of parsimonious assumptions on the underlying generative processes - for groups of time series.
We compare topological and thresholding network filtering approaches on both null model-based time-series ensembles and real data to test the long memory properties of multivariate financial time series. Each null model preserves different aspects of the time series, allowing to validate hypotheses about the long memory of market structures by ranking persistence decays of real time series against null models. 
We show how the edge and motif persistences - such as triangles and tetrahedra - of these models decay in TMFG-filtered graphs and graphs obtained by filtering correlation matrices through quantile thresholding. 
We show how topological filtering is a better suited tool to identify persistently correlated groups of securities throughout the market.
We compare TMFG with quantile thresholding of the correlation matrix, at fixed network density level observing that quantile thresholding yields analogous results to planar filtering for edge persistence, but it fails to identify motifs distributed throughout the market sample generating instead highly localised and clustered structures.
Further, we demonstrate that our findings have a practical application by introducing an unsupervised technique to identify groups of stocks which share strong fundamental price drivers. 
This technique can be of particular use in less traded markets, where identifying structures with shared fundamental price drivers might otherwise require in-depth knowledge of the companies.

The rest of the paper is structured as follows. Section \ref{method} describes the data, methods and definitions used for this work, Section \ref{results} outlines the main findings, Section \ref{analysis} discusses these findings and Section \ref{conclusion} concludes the work with suggestions for future works.

%The rest of the paper is structured as follows: Section ~(\ref{method}) describes the methods applied and defines measures which are used throughout the paper. Section ~(\ref{results}) describes the results obtained, with Section ~(\ref{results-long-term}) introducing long-term memory processes in persistence, Section ~(\ref{class_decay_exp}) analysing market development through its decay exponents, Section ~(\ref{sector_motifs}) illustrating the coherence of highly persistent motifs with sectors and Section ~(\ref{portfolio_applications}) outlining various results which highlight the importance of this work for portfolio allocation. Section ~(\ref{analysis}) then presents an analysis of the results from Section ~(\ref{results}) and Section ~(\ref{conclusion}) concludes the paper with a summary and thoughts for further work.

\section{\label{method}Materials and methods}

%In this section we describe the data used in this work, followed by a description of the time series null models and the filtering techniques applied to the correlation matrices. We then describe methods used to study the persistence of motifs and construct portfolios based on persistence.

\subsection{\label{data}Data}

We select the 100 most capitalised stocks from four stock markets:  NYSE, Italy, Germany and Israel's (400 stocks in total). 
Markets range from highly liquid and more developed ones such as the New York Stock Exchange and the Frankfurt Stock Exchange to less liquid markets such as the Italian Stock Exchange and the Tel Aviv Stock Exchange.

We investigate daily closing price data from Bloomberg for:

\begin{itemize}
  \item New York Stock Exchange (3/01/2014 - 31/12/2018);
  \item Frankfurt Stock Exchange (3/01/2014 - 28/12/2018);
  \item Borsa Italiana (Italian Stock Exchange) (3/01/2014 - 28/12/2018);
  \item Tel Aviv Stock Exchange (5/01/2014 - 1/1/2019).
\end{itemize}

The data respectively includes 1258 daily prices observations for the NYSE, 1272 for FSE and BI and 1225 for TASE.
%We though notice that some markets do not have sufficient/good quality data easily available, both in terms of number of stocks and price history. These are hence left for further work and we restrict ourselves to the abovementioned markets. 100 most capitalised stocks (at the moment of data collection) for the following markets: NYSE, Germany, Italy, Israel.

\subsection{\label{time_ser_null_mod}Time series null models}

We generate ensembles of null models which preserve an increasing number of properties of the real time series.

\paragraph{Random return shuffling}

Individual stock log-return ($r_t = \log Price_t - \log Price_{t-1}$) time series are randomly shuffled, i.e. a random permutation along the time dimension of each variable is applied, to obtain a null model for noise and spurious correlations. This model maintains the overall statistics of the values of each time series but eliminates any correlation structure.

\paragraph{Rolling univariate Gaussian generator}

We calculate the rolling mean $\mu_{t - \delta_t, t}$ and standard deviation $\sigma_{t - \delta_t, t}$ of the log-return series for each security separately. 
We then generate ensembles by sampling the return $r_t$ at each point in time from the (rolling) univariate Gaussian distributions with sample mean and standard deviation $r_t \sim \mathcal{N}(\mu_{t - \delta_t, t},\,\sigma_{t - \delta_t, t}^{2})$, with $\mathcal{N}(\mu,\,\sigma^{2})$ being a normal distribution with mean $\mu$ and standard deviation $\sigma$.
This intends to simulate the process as a simple moving average with uncorrelated time-varying Gaussian random noise.

%\paragraph{Maximum Entropy approach to multivariate time series randomization}

%After obtaining the log-return time series, we generate an ensemble through the method described in \cite{marcaccioli2019statistical}. 
%This is a maximum entropy method attempting to preserve sum, number of positive and number of negative values of rows and columns (hence with some concept of relation between time series) in a multivariate time series.
%This model improves upon the univariate rolling Gaussian as some relational characteristics are preserved through the summation constraints of columns - a given point in time common to all time series - and long-range dependence of time series is partially accounted for through constraints on rows (individual time series).

\paragraph{Stable multivariate Gaussian generator}

We calculate the mean $\boldsymbol{\mu}$ (for each security) and covariance matrix $\boldsymbol{\Sigma}$  throughout the whole length of the log-return time series. %between the series throughout their length.  
We then generate ensembles by sampling the vector of returns $\boldsymbol{r}_t$ at each point in time for all securities from the fixed multivariate Gaussian with empirical means and covariance matrix, $\boldsymbol{r}_t \sim \mathcal{N}(\boldsymbol{\mu},\,\boldsymbol{\Sigma}^{2})$. 
This intends to represent an underlying fixed market structure with sampling noise.

\paragraph{Rolling multivariate Gaussian generator}

After obtaining the log-return time series, we calculate the rolling mean $\boldsymbol{\mu}_{[t - \delta_t, t]}$ (for each security) and covariance matrix $\boldsymbol{\Sigma}_{[t - \delta_t, t]}$ between the series. 
We generate ensembles by sampling the return at each point in time $\boldsymbol{r}_t$ for all securities from the (rolling) multivariate Gaussian distributions with sample means and covariance matrices $\boldsymbol{r}_t = \mathcal{N}(\boldsymbol{\mu}_{[t - \delta_t, t]},\,\boldsymbol{\Sigma}_{[t - \delta_t, t]}^{2})$.
This intends to detect the changing market structure and simulate the process as being generated by a multivariate Gaussian distribution with time-varying constraints on structural relations.

\subsection{Correlation matrix estimation}

We then compute for the time series correlation matrices with exponential smoothing from rolling windows of $\delta=126$ trading days with smoothing factor of $\theta=46$ days. This is done for all realisations of each null model ensemble and for the real data.

Correlations are noisy measures of co-movements of financial asset prices, which are often non-stationary within the observation window. 
Longer time windows benefit the measure's stability, as we have more observations to estimate the $N (N-1)/2$ parameters of the matrix of $N$ assets. 
However, a longer observation window can come with the disadvantage of weighting more and less recent co-movements equally with the risk of averaging over a period in which the values are non-stationary. 
In order to compensate for this effect, we apply the exponential smoothing method for Kendall correlations ~\cite{pozzi2012exponential}. 
This allows for more stable correlations, as the method applies an exponential weighting to the correlation window, prioritising more recently observed co-movements. 
%Here we use Kendall correlations with exponential smoothing as per \cite{pozzi2012exponential}.

\subsection{Filtering: quantile thresholding and TMFG}

%In the present paper correlation matrices are obtained from Kendall correlations with exponential smoothing, applying the method by Pozzi et al. ~\cite{pozzi2012exponential}.

We apply two filtering techniques with fundamental differences. The first filtering method is quantile thresholding, which corresponds to hard thresholding to generate an adjacency matrix through the binarisation of individual correlations. For a correlation value $v_q$ corresponding to the quantile level $q$ of the matrix values, the adjacency matrix is defined as

\[   
A_{i,j} = 
     \begin{cases}
       \rho_{i,j} \geq v_q , & 1 \\
       \rho_{i,j} < v_q , & 0 .\\
     \end{cases}
\]

This filtering technique is entirely value-based with no structural or other constraints. 
We apply it by providing a quantile level $q$ which yields edge sparsity analogous to that of the corresponding TMFG filter.

The second filtering technique is the TMFG method ~\cite{massara2016network}. 
This topological filtering technique embeds the matrix with topological constraints on planarity in a graph composed by simplicial triangular and tetrahedral cliques. 
Edges are added in a constrained fashion with priority according to their (absolute) value. 
The graph essentially corresponds to tiling a surface of genus 0. 
This technique represents a filtering method that accounts for values, but also imposes an underlying chordal structural form which might help regularising the filtered graph also for probabilistic modeling \cite{PhysRevE.94.062306}. 
Furthermore, this technique imposes higher order structures, namely triangles and tetrahedra, which are known to be a feature of financial markets and social networks.

\subsection{Simplicial persistence}\label{simplicial_persistence_method}
%\paragraph{TMFG network motif persistence}
We focus on temporal persistence of tetrahedral and triangular simplicial complexes (motifs) in the TMFGs and graphs filtered via quantile thesholding constructed from correlations over rolling windows. 
TMFG networks can be viewed as trees of tetrahedral (maximal) cliques connected by triangular faces, these are triangular cliques with different meaning in the taxonomy, called separators. 
If removed, separators split the graph into two parts.
Not all triangular faces of the tetrahedral cliques are separators and we will refer to those which are not as triangles. % (these do not include separators in the way we shall refer to them). 

This distinction is discarded for the results in Section \ref{null_mod_res} in order to account for all triangles in the filtered graph, as quantile thresholding does not distinguish between triangular faces and separators.

\begin{figure}[!ht]
\caption{{\bf Motif persistence visualisation.}
Visual representation of a TMFG structure's motif (triangle) persistence in time. The green triangle in figure a) is persistent through figure c), while other two triangles (present in figure a) within the red triangle) do not persist due to the rewiring of an edge. Figure b) shows one of non-persistent triangles with dashed contour. The rewired edge is also dashed. This visualisation aims at showing the impact of edge rewiring on motif persistence and the difference between edge and motif persistence.}
\includegraphics[width=1\linewidth]{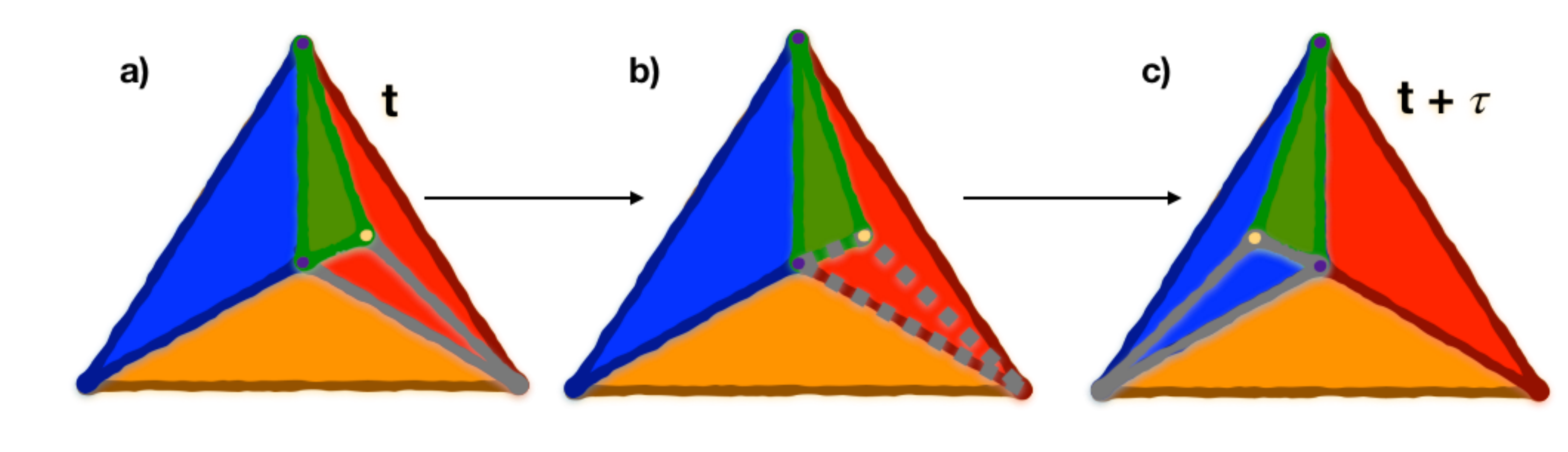}
\label{figure_TMFG_graph}
\end{figure}

A motif corresponding to clique $\mathcal{X}_c$ is considered soft-persistent at time $t + \tau$ if and only if the motif is present at both the initial time $t$ and at $t + \tau$. A visual intuition for motif (triangle) persistence through time is provided in Figure \ref{figure_TMFG_graph}.

We investigate the decay in the number of persistent motifs between filtered correlation networks with observation windows progressively shifted by one trading day and we quantify how the average persistence %$\langle P_{m}(\mathcal{X}^{\tau}) \rangle_{T, C}$ from Equation ~(\ref{eq_persist_motif_avg}) 
decays with the time shift $\tau$.

Here we use a form of soft persistence which is different from hard persistence (survival) of motifs which is more common in the literature ~\cite{dessi2018supernoder, musmeci2014risk}.
Specifically, the average motif persistence in the plateau regime is defined as
\begin{eqnarray}
\label{eq_persist_time_avg}
\centering
    \langle P_{m}(\mathcal{X}_c) \rangle_{T, \mathcal{T}} = \frac{1}{T} \cdot \frac{1}{\mathcal{T} - \tau_{plat}} \cdot \sum_{t = 0}^{T} \sum_{\tau = \tau_{plat}}^{\mathcal{T}} P_{m}(\mathcal{X}_c^{t, t+\tau}),
\end{eqnarray}
where $\tau_{plat}$ denotes the transition point to the plateau region.
The average persistence for the entire clique set over $T$ starting points at time shift $\tau$ is defined as
\begin{eqnarray}
\label{eq_persist_motif_avg}
\centering
    \langle P_{m}(\mathcal{X}^{\tau}) \rangle_{T, C} = \frac{1}{T} \cdot \frac{1}{|C|} \cdot \sum_{t = 0}^{T} \sum_{c \in C} P_{m}(\mathcal{X}_c^{t, t+\tau}).
\end{eqnarray}
Where, considering the motif sets $\mathcal{X}_C^{t} = \{ \mathcal{X}_i^{t} \}_{i = 1, ..., C}$ and $\mathcal{X}_C^{t + \tau} = \{ \mathcal{X}_i^{t + \tau} \}_{i = 1, ..., C}$, the binary persistence value of motif $c \in C$ at time $t$ and $t + \tau$ is

\begin{eqnarray}
\label{eq_defin_persist}
\centering
    P_{m}(\mathcal{X}_c^{t, t+\tau}) = (\mathcal{X}_c \in \mathcal{X}_C^{t}) \land (\mathcal{X}_c \in \mathcal{X}_C^{t + \tau})
\end{eqnarray}

%Where $P_{m}(\mathcal{X}_c^{t, t+\tau})$ represents the binary persistence value of motif $c \in C$ at times $t$ and $t + \tau$.

We obtain the power law fit for the decay law and identify two regimes: one with a faster decay followed by one with a slower decay. 
The transition point $\tau_{plat}$ is computed by minimising the unweighted average mean squared error (MSE) between the two fits over all possible transition points in time.

We also compare the decay exponents for multiple random stock selections over different markets to identify whether the steepness of motif decay (edge, closed triad or tetrahedron clique) is indicative of market stability/development stage.
We further investigate more liquid markets such as the NYSE from both a quantitative and qualitative point of view.% as follows.
We classify motifs in the plateau by their soft persistence and study the sector structure of the most persistent motifs. 
%We also verify that these motifs are not trivially retrieved by maximum correlation edges or motifs in the correlation matrix.

In order to further justify the analysis of motifs over individual edges, we test the null hypothesis that motifs are formed by edges in the network whose existence is not mutually dependent. 
The assumption would imply that coexistence of edges in motifs is not statistically significant and that motif structures have no extra persistence beyond the individual edges that form them. 
The hypothesis being tested implies that motif persistence is simply the result of persistence characterising their component edges:
%This is falsified by the consistently lower decay exponent (in modulus) for adjusted persistence of triangular motifs.

\begin{eqnarray}
\label{eq_motif_meaning}
\centering
    P_{m}(\boldsymbol{\chi}_c^{t,t+ \tau}) = P_{m}(\boldsymbol{\chi}_{c1}^{t,t + \tau}) \cdot P_{m}(\boldsymbol{\chi}_{c1}^{t,t + \tau}) \cdot P_{m}(\boldsymbol{\chi}_{c3}^{t,t + \tau}),
\end{eqnarray}

where the motif and its edges are defined as $\boldsymbol{\chi}_c^{t,t + \tau} = \{ \boldsymbol{\chi}_{c1}^{t,t + \tau} , \boldsymbol{\chi}_{c2}^{t,t + \tau} , \boldsymbol{\chi}_{c3}^{t,t + \tau} \}$.

%\subsection{Portfolio construction}

%SHOULD WE MOVE THIS TO THE PREVIOUS SECTION AND REMOVE THIS ONE?

In order to provide an application to systemic risk, we construct a portfolio containing all stocks in the ten most persistent motifs in the plateau region, as defined in Equation \ref{eq_persist_time_avg} (for each market). 
We then compare its volatility with that of random portfolios with the same number of assets. 
%We also construct a portfolio containing all stocks in the ten most persistent motifs and compare its volatility with that of portfolios containing only part of the motifs (breaking the persistent motif structure).

%We conclude by defining the persistence measure $P_m(v_i)$ in Equation ~(\ref{eq_portfolio_measure}) to compare random portfolios weighted by $1/ \sigma$ with those weighted by $1/ P_m(v_i)$. We do this for the four different markets, with all results showing meaningful volatility reductions.

%The measure presented in Equation ~(\ref{eq_portfolio_measure}) is defined for each vertex $v_i$ in the network as the sum over all $\langle P_{m}(\mathcal{X}_c) \rangle_{T, \mathcal{T}}$ (average pesistence of motif $\mathcal{X}_c$ in the plateau) where vertex $v_i$ belongs to clique $\mathcal{X}_c$. %, with  the vertex of the network associated with the measure.

%in the TMFG network obtained from the correlation matrix (Kendall with exponential smoothing)

\section{\label{results}Results}

The main findings of this work are described in this section, starting with an overview of results on the long memory of edges and simplicial complexes in TMFG-filtered correlation networks. 
The section continues with an analysis of null models of financial market structures, described in Section \ref{time_ser_null_mod}, and a comparison with real data to gain insights about the generative process of the stochastic structure. 
We then suggests how soft persistence captures the underlying change in market structure by relating its decay exponent to the stage of development (a proxy for stability) or average traded volume in the market (a proxy for liquidity which yields well-defined stable structures). %We investigate systemic risk by showing that the most persistent motifs correspond to stocks in the same sector. 
We conclude the section with results in systemic risk applications to financial portfolios where we show that the most persistent motifs correspond to stocks in the same sector and demonstrate how the portfolio of 10 most persistent motifs is highly volatile and systemic.
%The main findings of this work are described in this section, with Section \ref{results-long-term} discussing the long memory of edges and motif structures in TMFG financial correlation networks. Section \ref{null_mod_res} deals with assumptions about the evolving structure of financial markets and how they compare to observations of edge and motif persistence in the null models described in Section \ref{time_ser_null_mod}. Section \ref{class_decay_exp} suggests how ``soft'' persistence captures the underlying change in market structure by relating its decay exponent to the stage of development (a proxy for stability) or average traded volume in the market (a proxy for liquidity which yields well-defined stable structures). Section \ref{sector_motifs} shows how the most persistent motifs correspond to stocks in the same sector. Section \ref{portfolio_applications} present the results in systemic risk applications to financial portfolios, demonstrating how the portfolio of 10 most persistent motifs is highly volatile and systemic.

\subsection{\label{results-long-term}Long-term memory of motif structures}

\begin{figure}[!ht]
\caption{{\bf Persistence Decay.}
Decay of triangular clique faces, separators and clique motifs persistence for 100 NYSE stocks, as a function of time interval $\delta_t = [0, 900]$ (average over 200 starting points). 
The two power-law regimes are identified by the minimum MSE sum of the fits.}
\includegraphics[width=1\linewidth]{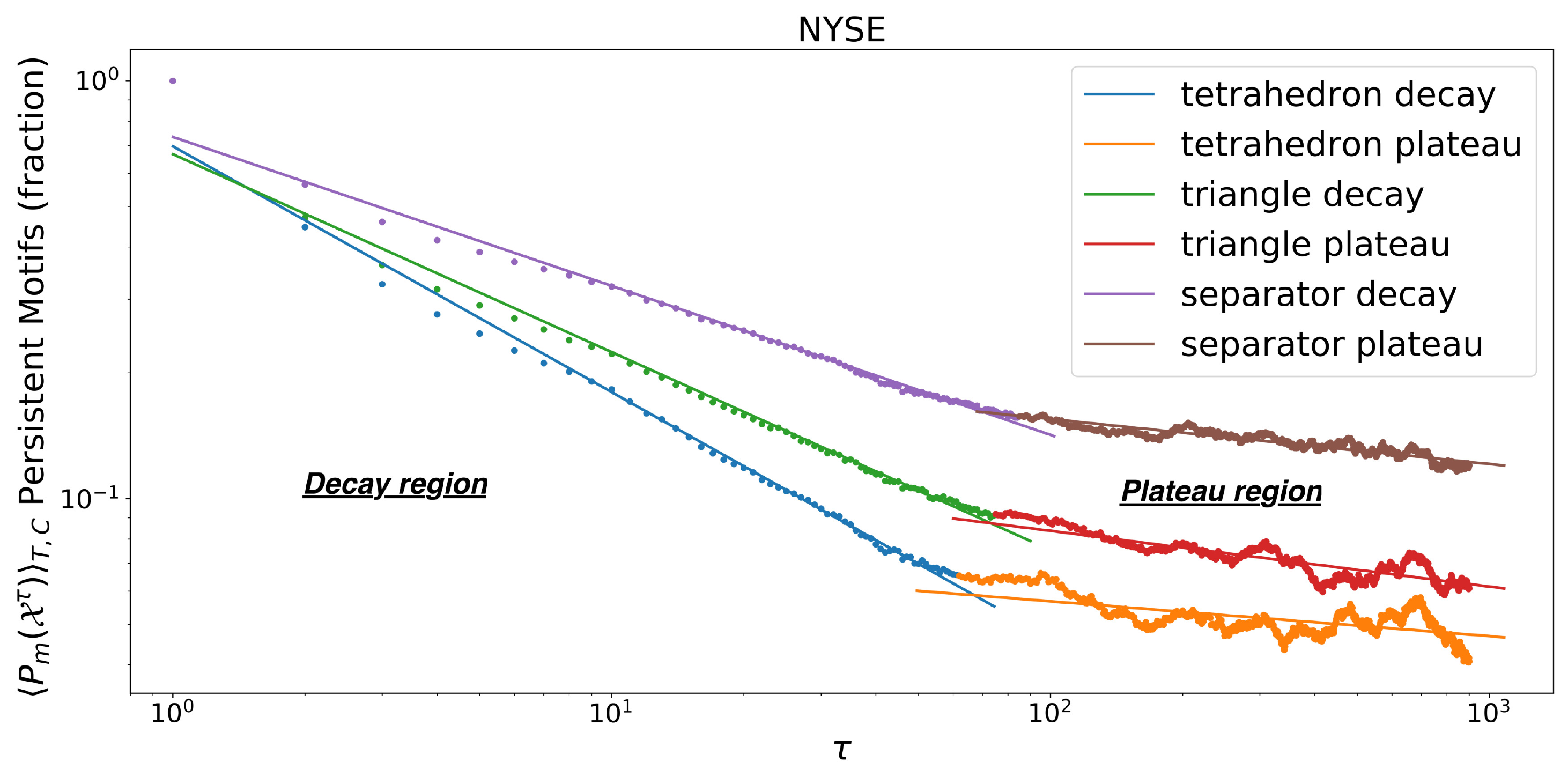}
\label{figure2}
\end{figure}

%\begin{figure}[!htb]
%\centering
%\includegraphics[width=1\linewidth]{power_law_decay_NYSE.pdf}
%\caption{Decay of triangular clique faces, separators and clique motifs overlap between layers for 100 NYSE stocks, as a function of time shift $\tau = [0, 900]$ (average over 200 values of $t$). The two power-law regimes are identified by the minimum MSE sum of the fits.}
%\label{figure1}       % Give a unique label
%\vspace{-0.4cm}
%\end{figure}
%in the number of ``soft'' persistent motifs
The plot in Figure ~(\ref{figure2}) shows the power law decay (evident from the linear trend in log-log scale) in $\langle P_{m}(\mathcal{X}^{\tau}) \rangle_{T = 200, C}$ vs. $\tau$, followed by a plateau region that also decays as a power law, but with a smaller exponent. We also observe that all motif decays have $\tau_{plat} \in [\delta t_{window}/2, \delta t_{window}]$, where $\delta t_{window}$ represents the length of the estimation window of the correlation matrix. The window used has $\delta t_{window} = 126$ trading days and a value of $\theta = 46$ for exponential smoothing, as per ~\cite{pozzi2012exponential}. 
The choice of $\delta t_{window}$ corresponds to roughly 6 months of trading and satisfies $N < \delta t_{window}$, with $N$ the number of assets in the correlation matrix. 
The correlation matrix is hence well-conditioned and invertible.
On the other hand the exponential smoothing with $\theta = 46$ mainly considers recent observations from the latest few months.

There are $N - 3 = 97$ tetrahedral cliques in the starting TMFG networks and $3N - 8 = 292$ face triangles.

%The plot in Figure \ref{figure2} is analogous to that in Figure \ref{figure1}, but with the number of persistent motifs scaled by the total number of motifs in an $N$-nodes TMFG. This corresponds to the number of motifs in the first temporal layer.
%Figure \ref{figure2} provides a better visual comparison of decay rates, with tetrahedral cliques having a higher decay rate, as they are composed of more edges (six) which need to persist simultaneously.

In Figure ~(\ref{figure2}) we notice that the minimum MSE for the two linear fits is achieved at the transition point between the decay phase and the plateau. 
The transition point $\tau_{plat}$ can therefore be identified by minimising a standard fit measure with two phases, which strengthens the unsupervised nature of our method. 
The method for minimum MSE search is described Section \ref{simplicial_persistence_method}.
%Section ~(\ref{method}).

\subsection{\label{null_mod_res}Null models of persistence in filtered structures}

We report results for the edge and motif (triangle) persistence for real data as well as for the null models described in Section \ref{time_ser_null_mod}. 
We compare real data with null models and TMFG filtering with quantile thresholding.

\begin{figure}[!ht]
\caption{{\bf Edge persistence decay of null models.}
Edge persistence decay with $\delta_{\tau}$ for the time series null models of market returns and real data for the NYSE. We notice how for both TMFG filtering and quantile thresholding the real data lies between the rolling multivariate Gaussian ensemble and the stable multivariate Gaussian ensemble. 
This indicates that the real market structure does evolve slowly in time, but with persistence beyond what can be inferred from estimates of its covariance structure.}
\includegraphics[width=1\linewidth]{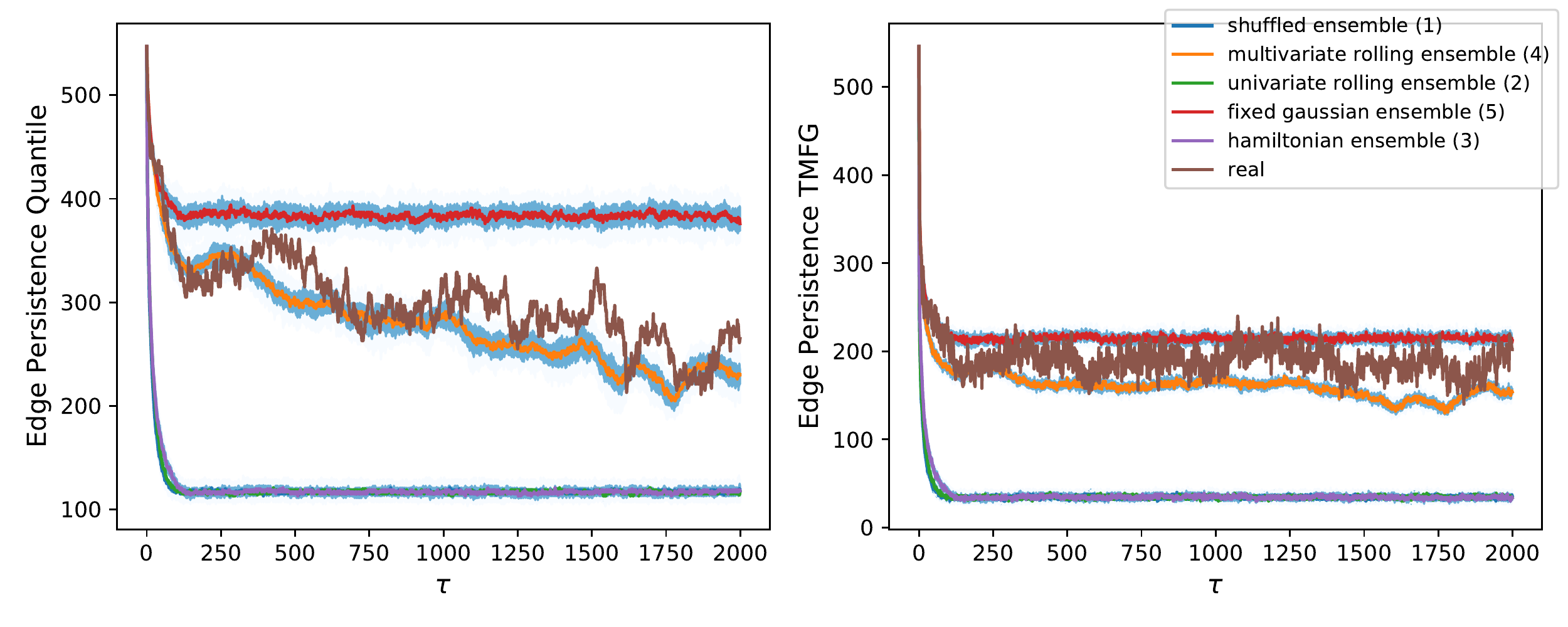}
\label{figure4}
\end{figure}

Figure \ref{figure4} shows the decay in edge persistence for both filtering methods. We notice that the random shuffling null model lies at the bottom, as it should produce completely random structures with little residual persistance due to probabilistic combinatorics and structural filtering constraints in the TMFG. 
This shows that persistence is not an artifact of any of the filtering techniques used and not a mere result of return volatility of individual assets (which is preserved by return shuffling). 
From Fig. ~(\ref{figure2}) we also notice that the rolling univariate Gaussian model lies just above as it does not account for structure at all and only preserves rolling means and standard deviations, this shows how persistence cannot merely be attributed to common long term trends or volatility variations.
This null model carries some broad sense of structure and market direction and it shows how persistence does not merely originate from overall market trends.
%The two null models described above lie significantly below the two others, as they do not explicitly account for individual relations between stocks. 
We then find a second cluster, of structured models, with the rolling multivariate Gaussian at the bottom. 
This shows how market persistence goes beyond asset means and covariance, even after spurious structures have been removed.
We then find the real data, just below the stable multivariate Gaussian. This shows how markets have slowly evolving structures. % less persistent than those generated by a single underlying matrix.

\begin{figure}[!ht]
\caption{{\bf Motif (triangle) persistence decay of null models.}
Motif persistence decay with $\delta_{\tau}$ for the time series null models of market returns and real data for the NYSE. We notice how for TMFG filtering the real data still lies between the rolling multivariate Gaussian ensemble and the stable multivariate Gaussian ensemble (as in Figure \ref{figure4}). 
We instead notice that the decay ordering is noisier for quantile thresholding, showing how the method's focus on individual connections affects it generalisation to motifs. 
This is despite the higher number of motifs in the quantile thresholding graph.}
\includegraphics[width=1\linewidth]{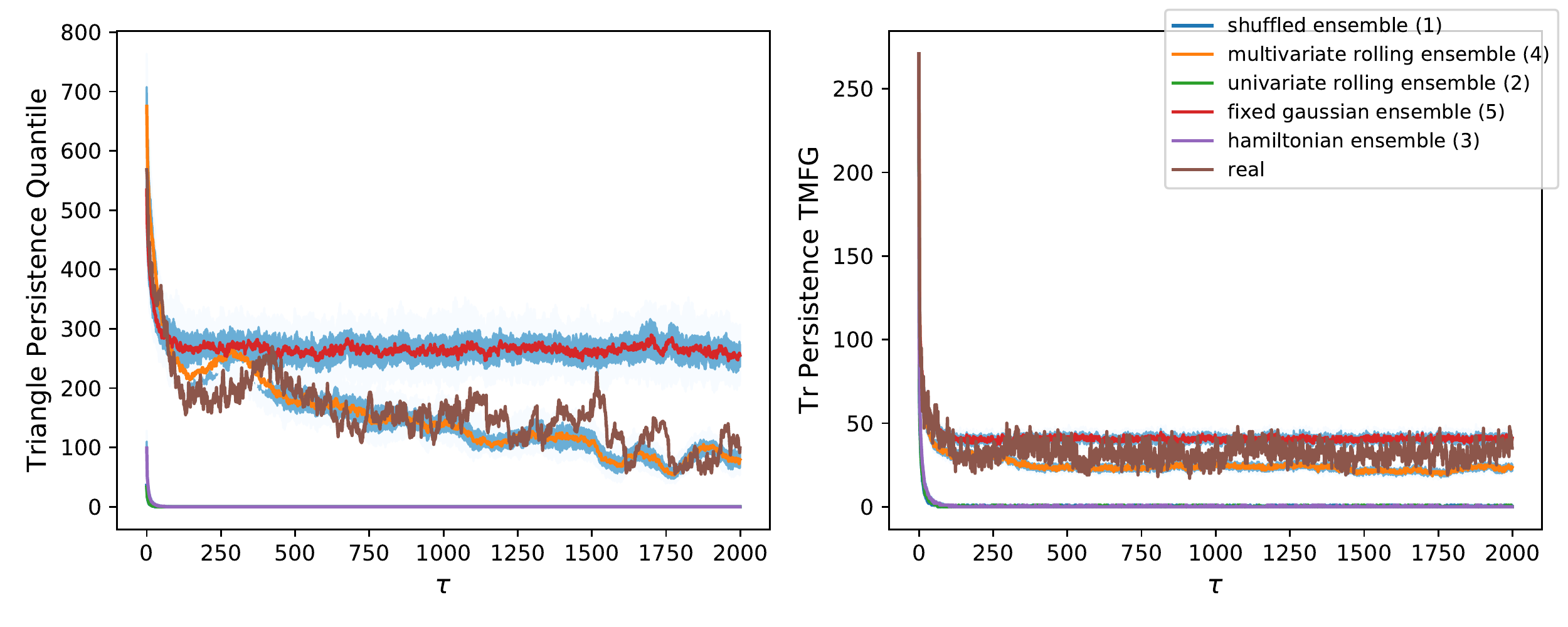}
\label{figure5}
\end{figure}

Figure \ref{figure5} shows the decay in triangular motif persistence for both filtering methods. 
We notice results analogous to those in Figure \ref{figure4} for TMFG filtered graphs. Graphs filtered through quantile thresholding instead show a high level of noise in their top cluster (where structure is present). 
A higher number of motifs than those of the TMFG is found, but the ranking of null models is at times inconsistent, as well as the position of the decay curve for real data. 
%We investigate this further in Figure \ref{figure6} where we notice a stable clustering coefficient for quantile thresholding, whilst the TMFG sees a decay. 
We would have expected some triangles to break when looking at edge persistence only, as well as to find that the clustering coefficient decreases in persistent graphs (as it does in TMFG graphs). 
The clustering coefficient for quantile thresholding-persistent graphs is also found to be much higher, suggesting that the filtered structure is highly localised and clustered, while that of the TMFG is more distributed, identifying systemic groups of stocks throughout the market structure.

\subsection{\label{class_decay_exp}Market classification via decay exponent}
 
We now consider how the decay exponent of TMFG graphs behaves across markets. 
Table ~(\ref{tab1}) compares the decay exponents for cliques, triangular motifs and clique separators in the NYSE, German stock market, Italian stock market and Israeli stock market. 
The decay exponent $\alpha$ is obtained from the fit based on the following expression,

\begin{eqnarray}
\label{eq_persist_decay_fit}
\centering
    \langle P_{m}(\mathcal{X}^{\tau}) \rangle_{T, C} = \beta \cdot \tau^{\alpha}
\end{eqnarray}

%\frac{1}{T} \cdot \frac{1}{|C|} \cdot \sum_{t = 0}^{T} \sum_{c \in C} P_{m}(\mathcal{X}_c^{t, t+\tau})

\begin{table}[!ht]
%\begin{adjustwidth}{-2.25in}{0in} % Comment out/remove adjustwidth environment if table fits in text column.
\centering
\caption{
{ \bf Exponents for the decay power law regime computed with MSE.
The analysis refers to 100 randomly selected stocks amongst the 500 most capitalised, over time intervals $\tau = \left[0, 900 \right[$ and $t = \left[ 0, ..., 200 \right[$ different initial temporal network layers. 
For all motif analyses in this work, triangles and separators constitute non-overlapping sets, as these represent theoretically and taxonomically different structures and decay characteristics.}}
%\begin{ruledtabular}
\begin{tabular}{llll}

Market & Clique & Triangular Motif & Clique Separator\\ \hline
%\multicolumn{4}{|l|}{\bf Heading1} & \multicolumn{4}{|l|}{\bf Heading2}\\ \thickhline
NYSE & -0.392 & -0.493  & -0.245\\ 
Germany & -0.792 & -0.598  & -0.381\\ 
Italy & -0.785 & -0.811  & -0.174*\\ 
Israel & -1.024 & -0.866  & -0.728\\ 

\end{tabular}
%\end{ruledtabular}
\begin{flushleft} * Result compromised by regimes not well identified for motif decay in large systems ($\approx 100$ stocks).
\end{flushleft}
\label{tab1}
%\end{adjustwidth}
\end{table}

We notice from the results in Table ~(\ref{tab1}) that the NYSE, which is clearly the most developed and liquid stock market, has the lowest decay exponent (in modulus, which corresponds to the slowest decay) for both cliques and triangles. This indicates that its correlations are more stable on a shorter time window.% due to a higher signal to noise ratio. 
Germany and Italy have similar values for clique exponents, with Germany seemingly more stable in terms of triangular motifs. 
Israel, a younger and less liquid stock market, follows with a faster decay in both tetrahedral cliques and triangular motifs. 
The ordering of these markets is not clearly identifiable in clique separators as noise in the data does not allow for the two decay regimes to be correctly identified in all markets (in this case for Italy). 
Separators have a distinct role and meaning in the graph's taxonomy and further work should allow for a more thorough analysis of those.

We observe promising results for a monotically increasing relation between the decay exponent and the average daily volume of the market. The solidity of this result shall be investigated in future works.
%Section ~(\ref{results-long-term})

%Tetrahedral cliques can be viewed as groups of stocks representing maximal cliques (of four vertices) in the graph, while triangles are smaller cliques (three-cliques) which constitute the faces of the tetrahedral maximal cliques.
In Table ~(\ref{tab1}) the decay exponent is not adjusted by the probability that all edges in the clique must be present in the temporal layer for the clique to exist. 
We show in Table ~(\ref{tab2}) that, when adjusted by the probability of all its edges existing simultaneously, triangular motifs have a slower decay than individual edges. 
The results in Table ~(\ref{tab2}) are obtained from a set of randomly selected stocks different to those used for Table ~(\ref{tab1}). This adds further confidence in the results and their generality.

We stress that Table ~(\ref{tab2}) falsifies the hypothesis that motifs are formed by edges in the network whose existence is not mutually dependent (Equation ~\ref{eq_motif_meaning}). 
This is falsified by the consistently lower decay exponent (in modulus) for adjusted persistence of triangular motifs. 
We can then conclude that motifs are more stable structures across temporal layers of the network, with significant interdependencies in their edges' existence.
%Section ~(\ref{method})

\begin{table}[!ht]
%\begin{adjustwidth}{-2.25in}{0in} % Comment out/remove adjustwidth environment if table fits in text column.
\centering
\caption{
{\bf Exponent for the power law decay regime identified by MSE in different sample markets. 
The analysis refers to 100 randomly selected stocks amongst the 500 most capitalised, over time intervals $\tau = \left[0, 900 \right[$ and $t = \left[ 0, ..., 200 \right[$ different initial temporal network layers.}}
%\begin{ruledtabular}
\begin{tabular}{llll}

Market & Edge & Triangular Motif & Triangular Motif**\\ \hline
%\multicolumn{4}{|l|}{\bf Heading1} & \multicolumn{4}{|l|}{\bf Heading2}\\ \thickhline

NYSE & -0.164 & -0.398 & -0.133\\ 
Germany & -0.265 & -0.471 & -0.157\\ 
Italy & -0.144* & -0.458 & -0.153\\ 
Israel & -0.397  & -0.830 & -0.277\\ 

\end{tabular}
%\end{ruledtabular}
\begin{flushleft} * Result compromised by regimes not well identified for edge decay in large systems ($\approx 100$ stocks)

** Motif exponent adjusted by the probability of simultaneous edge persistence in the motif).
\end{flushleft}
\label{tab2}
%\end{adjustwidth}
\end{table}

\subsection{\label{sector_motifs}Sector analysis in persistent motifs}

\begin{figure}
\caption{{\bf Persistent NYSE motifs visualised.}
Network representation of the ten most persistent triangular motifs in the TMFG layers for the 100 most capitalised stocks of the NYSE..}
\includegraphics[width=1\linewidth]{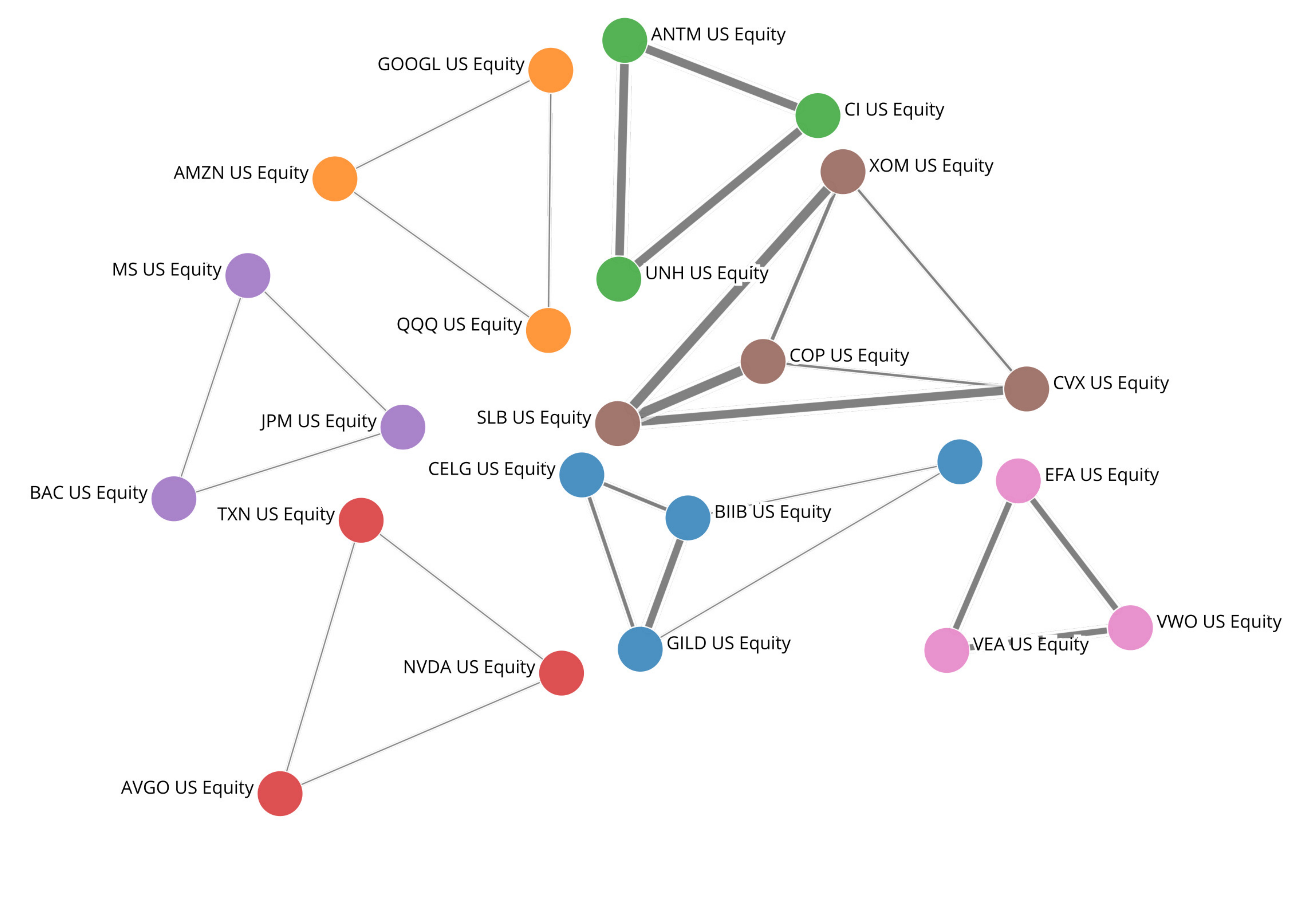}
\label{figure3}
\end{figure}

Figure ~(\ref{figure3}) provides a visualisation of the network components formed by the ten most persistent triangles in the NYSE. We observe that all strongly persistent triangles have elements which belong to the same industry sector. Table \ref{tab3} shows this for the same ten triangles displayed in Figure ~(\ref{figure3}).
%an illustration of this for the triangles in Figure \ref{figure3} is provided in Table \ref{tab3}.
%Section ~(\ref{intro})
We notice that stock prices in the sectors in Table ~(\ref{tab3}) are mostly driven by sector-wide fundamentals, which justify the persistent structure in the long term. Other motifs are constituted by ETFs and their main holdings \footnote{The reason for the existence of these motifs is intuitive and does not affect our analysis, as ETF-related motifs are unlikely to be present in the network formed by a random selection of stocks or by stocks in a portfolio. These motifs are present here as we focus on the 100 most capitalised securities in the NYSE, which include ETFs.}.

\begin{table*}
%\begin{adjustwidth}{-2.25in}{0in} % Comment out/remove adjustwidth environment if table fits in text column.
\centering
\caption{
{\bf Motif components and Financial Times sector affiliation for the ten most persistent motifs in the NYSE's 100 most capitalised stocks.}}
%\begin{ruledtabular}
\begin{tabular}{llllll}
%\hline
Security 1 & Security 2 & Security 3 & FT Sector\\ \hline
%\multicolumn{4}{|l|}{\bf Heading1} & \multicolumn{4}{|l|}{\bf Heading2}\\ \thickhline
Biogen Inc & Gilead Sciences Inc & Celgene Corp & Biopharmaceutical\\ \hline

UnitedHealth Group Inc & Cigna Corp & Anthem Inc & Health Care\\ \hline

Biogen Inc & Gilead Sciences Inc & Amgen Inc & Biopharma/tech\\ \hline

Bank of America Corp & JPMorgan Chase \& Co & Morgan Stanley & Financials-Banks\\ \hline

Vanguard FTSE ETF** & MSCI EAFE ETF & Vanguard FTSE ETF*** & Index ETFs\\ \hline

Invesco QQQ Trust* & Amazon.com Inc & Alphabet Inc & Tech\\ \hline

ConocoPhillips & Schlumberger NV & Exxon Mobil Corp & Oil \& Gas\\ \hline

NVIDIA Corp & Texas Instruments Inc & Broadcom Inc & Tech Hardware\\ \hline

Chevron Corp & Schlumberger NV & Exxon Mobil Corp & Oil \& Gas\\ \hline

Chevron Corp & ConocoPhillips & Schlumberger NV & Oil \& Gas\\ \hline

\end{tabular}
%\end{ruledtabular}
\begin{flushleft} * ETF on NASDAQ - Top Holdings include Amazon, Facebook, Apple, Alphabet

** Vanguard FTSE Developed Markets Index Fund ETF Shares

*** Vanguard FTSE Emerging Markets Index Fund ETF Shares
\end{flushleft}
\label{tab3}
%\end{adjustwidth}
\end{table*}

We also investigate whether motif persistence and motif structures can be easily retrieved from the original correlation matrix. 
The purpose of this is to check that our TMFG filtering method is not redundant and trivially replaceable.
To test this, we consider the ten most present persistent triangles across the plateau region and check their overlap with the ten most correlated triplets in each unfiltered correlation matrix. 
We find that no more than one triangle lies in the intersection between the two sets, in each temporal layer. %corresponds in each layer.
We also check the correlation between motif persistence and the average sum or product (results are equivalent for our purpose) of its individual edges' correlation for all unfiltered correlation layers. 
We observed through the Pearson and Kendall correlation values that the two measures are only loosely related, as correlation explained no more than $20\%$ of the variance in the set of variables with large persistence. % values. 
%This result is particularly significant for a wide power law distribution as that of persistence values. 

%Figure \ref{figure_trivial_corr} shows a scatter plot demonstrating that the measure is not trivially retrievable from the correlation matrix.

\subsection{\label{portfolio_applications}Portfolio volatility and systemic risk of persistent motifs vs. random portfolios}\label{portfolio_random_res}

%\begin{figure}[!ht]
%\includegraphics[width=1\linewidth]{portfolio_volatility_10_triangles_random_portfolios_all}
%\caption{{\bf Random vs. persistent portfolios volatility.}
Portfolio volatility distribution for the 100 most capitalised stocks in %the NYSE (a), German stock market (b), Italian stock market (c) and Israeli stock market (d). 
%The reference portfolio (red bar) contains all stocks in the 10 most persistent triangles and distribution portfolios are formed from a random selection of stocks (mean distribution volatility represented by the green bar). (LABELS 4,5,6,7)}
%\label{fig_random_portfolios}
%\end{figure}

We check that a portfolio formed by the 10 most persistent motifs in each market has a highly enhanced out of sample volatility due to its stable correlations.

To do this, we consider the volatility of the motif portfolio and a distribution of volatilities for $10^5$ randomly selected portfolios with the same number of stocks.

As expected, we observe the motif portfolio to yield a volatility $vol_{motif}$ close to the higher end of the distribution, i.e. $( vol_{motif} - \langle \mathbf{vol_{random}} \rangle) > 2 \cdot \sigma(\mathbf{vol_{random}})$, throughout the considered markets. We should highlight that the volatility of portfolios is evaluated out of sample with respect to the period the persistence was calculated on, showing that this method is not only observational, but also predictive.

Due to the more theoretical nature of this work, we refer the interested reader the work by some of the authors of this paper for a more thorough analysis of portfolio applications and forecasting \cite{turiel2019sector}.

\section{\label{analysis}Discussion}

The power law decay of edge and simplicial soft persistences reported in Figure~\ref{figure2} suggests that market structures are characterised by a slow evolution which allows for long memory in temporal layers. 
This decay type is in contrast with an exponential decay of the persistence which would imply instead short or no memory in the system.
This observation is in line with the works by Bouchaud et al. and Lillo at al. in~\cite{bouchaud2004fluctuations, lillo2004long, bouchaud2009markets, di2005long}, where power law decays in autocorrelation are identified as manifestations of long-memory processes in efficient markets. 
However, it extends the concept to higher order structures. %Power law decays are meaningful as they are slower than exponential decays, which would represent processes with no long-term memory.

The comparison between soft persistence in correlation structures from real data and artificial data generated from different null models (Figs. \ref{figure4} and \ref{figure5}) demonstrates that the  persistence of real structures goes beyond all univariate null models, hence confirming long memory as a characteristic requiring structural constraints. Also we demonstrate that real structures overcome the persistence of the rolling multivariate Gaussian, hence suggesting that pairwise covariances and moving averages do not suffice to induce the long memory present in real markets. As per the analysis on motif persistence beyond those of individual edges, we suggest that higher order relations in terms of structural evolution are present.
The ordering of null models in Figure \ref{figure4} further supports the validity of the persistence measure.

The comparison of simplicial persistence of triangles between quantile thresholding and TMFG filteres graphs, reported in Figure \ref{figure5}, reveals that quantile thresholding struggles to separate the decay of real structures from that of rolling Gaussian generated ones. This could be attributed to the ``local'' nature of the method, which matches the pairwise interpretation of relations in generating from a rolling Gaussian. TMFG filtered graphs instead, perhaps due to their non-local embedding, provide a consistent ordering of null models with relatively low noise.

The ability to correctly identify persistent motifs throughout the market sample is essential as the most persistent motifs were found to be highly systemic (Section \ref{portfolio_random_res}).
Persistent structures in quantile thresholded graphs present higher and more stable clustering coefficients. This suggests a very localised and compact structure. TMFG filtered graphs instead present a lower clustering coefficient and a decay with $\tau$, as expected since some structures break. This is further evidence of the ability of the TMFG filtering method to identify meaningful persistent structures throughout the market. 
The issue with quantile thresholding is likely due to the method being merely value-based with no sensible structural constraint, differently from the TMFG.

%Major works in quantitative finance study the autocorrelation of returns and the long memory in return volatility autocorrelation. This work adds the missing piece of analysing autocorrelation of market structures, through persistence of its filtered correlation matrix.

The ranking of national markets based on their decay exponents in Table~\ref{tab1} can be interpreted in terms of the reduction of estimation noise in more liquid markets, as large deviations become less likely and correlations as well as prices more reflective of the underlying generative processes and structures. Structures are perhaps clearer too and deviations are exploited more quickly if they emerge.
This suggests that more efficient and capitalised markets are characterised by structures which are more stable in time and better reflected by the data. 
The decay exponent ranking also leads to the conclusion that more developed markets are characterised by more meaningful underlying structures and cliques, suggesting that systemic risk may represent a greater threat in developed markets. 
%Furthermore, as per the discussion above, a slower decay indicates stronger long-term memory in the system.

%These are interesting observations in relation to the efficient market hypothesis, indicating that more liquid and efficient systems display more stable, autocorrelated and predictable market structures.

The results in Table~\ref{tab2} support the hypothesis that motifs constitute meaningful structures in markets, beyond their individual edges. These results test the independence null model of individual edges in motif formation and show solid evidence to reject it.
We can then conclude that highly persistent motifs are not a mere consequence of highly persistent individual edges, but also of the correlation in those edges existing concurrently. This results ties in with the above discussion on the issues with locality of filtering methods and generative processes.

Table~\ref{tab3} strengthens the importance of persistent motifs. 
Indeed, the ten most persistent motifs visualised in Figure~\ref{figure3} are representative of industry sectors in the NYSE. 
These sectors are not identified by the motifs with higher edge correlation, which instead are dominated by motifs often due to correlation noise in high volatility stocks. 
Persistence and the identification of persistent motifs are hence found to be non-trivial with respect to correlation strength of individual edges or motifs. 
The impact on portfolio diversification of the motifs in Figure~\ref{figure3} indicates that these structures are highly relevant for systemic risk and portfolio volatility, with high predictive power provided by the long memory property of persistence, which is an intrinsic temporal feature.
As these motifs are not characterised by noticeably strong correlations, a common variance optimisation of the portfolio is unlikely to optimise the weights to sufficiently minimise the risk from these highly systemic structures.
%Section~\ref{portfolio_applications}

The systemic relevance of persistent motifs as well as their out of sample forecasting power are shown by the results in Section \ref{portfolio_random_res} and in \cite{turiel2019sector}, where significantly higher out of sample portfolio volatility is observed for the portfolio of persistent motifs. The motif portfolio volatility is significantly above both the mean and median of the random portfolios' volatility distribution.

This is a first example of how just selecting stocks from the ten most persistent motifs forms a portfolio with higher long term volatility. Clearly when aiming for a reduction in systemic risk, low volatility (the opposite) is the objective. 
The observations from Section \ref{portfolio_random_res} and \cite{turiel2019sector} lay the ground for the construction of portfolios where asset weights aim to reduce the volatility originating from persistent correlations in motif structures.
%Section~\ref{portfolio_applications}
%This is done in the long-only portfolio diversification across markets subsection of the Results section where we propose a simple node-specific measure for portfolio weighting and selection. 
%We show the out of sample volatility distribution of random portfolios with weights optimised as $1/P_m(v_i)$ to be significantly lower than the distribution of portfolios optimised as $1/\sigma$, a widespread industry standard for portfolio weighting. 
%This result can be explained by the persistence in time being the base of this measure, providing strong out of sample predictive power. 
%Volatility is known to change in the medium to long term for most assets, whilst correlation is also difficult to estimate due to noise in the data and measures. 
%This result greatly demonstrates the applicability of this work for portfolio optimisation by providing a mapping from persistence-related observations to a direct measure of portfolio performance. 
%Future works should investigate a technique to jointly optimise portfolio weighting for both persistence and volatility.

\section{\label{conclusion}Conclusion}

The present work introduces the concept of simplicial persistence, focusing on the soft persistence in simplicial cliques. 
This measure is applied to a complex system with a slowly evolving stochastic structure, namely financial markets.
The graph structures are obtained from Kendall correlations with exponential smoothing and filtered with the TMFG or through quantile thresholding.
The slow evolution of these systems with time manifests long memory in their structure with a two regime power law decay in persistence with time. The transition point between regimes is identified in an unsupervised way with mean-squared error minimisation. 

Null models of market structure are then used to test hypotheses about the generative process underlying the system. Two persistence decay clusters are observed, where the least persistent corresponds to null models with no structural constraints and the upper one (most persistent) comprises the rolling multivariate Gaussian (lowest), real data, and the stable multivariate Gaussian (highest).

Simplicial persistence of higher order structures in real data and null models is hardly recognised by value-based thresholding methods which are unable to identify persistent cliques throughout the market sample.
Decay exponents for different markets are then observed to provide a ranking corresponding to their liquidity or stage of development, which suggests that, despite these systems being less predictable in their individual series, they are more stable and predictable in terms of structure.
Most persistent motifs are found to correspond to sectors where the price of stocks is mostly driven by sector-wide fundamentals.

Based on the ability of simplicial persistence to forecast and identify strongly correlated clusters of stocks, the impact of persistence-based systemic risk on portfolio volatility is verified with a comparison between the ten most persistent motifs portfolio and random portfolios of the same size \cite{turiel2019sector}.

The present work provides further evidence of how network analysis and complex systems can enhance our understanding of real world systems beyond traditional methods.
Our results and methods lay the ground for future studies and modelling of the evolution of stochastic structures with long memory.

\section{\label{acknowledgments}Acknowledgments}
TA and JT acknowledge the EC Horizon 2020 FIN-Tech project for partial support and useful opportunities for discussion. JT acknowledges support from EPSRC (EP/L015129/1). TA acknowledges support from ESRC (ES/K002309/1), EPSRC (EP/P031730/1) and EC (H2020-ICT-2018-2 825215).

\bibliography{apssamp}

\end{document}